\documentclass[journal=jacsat,manuscript=article]{achemso}

\usepackage{graphicx}
\usepackage{dcolumn}
\usepackage[table,xcdraw]{xcolor}
\usepackage{bm}
\usepackage{color}
\usepackage{amsmath}
\usepackage{mathtools}
\usepackage{soul}
\usepackage[normalem]{ulem} 
\usepackage[version=3]{mhchem} 

\newcommand{\new}{}
\newcommand{\stkout}[1]{}

\author{Antonia Kowalewski}
\affiliation[SFU]
{Department of Physics, Simon Fraser University, 8888 University Drive, Burnaby, British Columbia, V5A 1S6 Canada}
\author{Nancy R. Forde}
\email{nforde@sfu.ca}
\affiliation[SFU]
{Department of Physics, Simon Fraser University, 8888 University Drive, Burnaby, British Columbia, V5A 1S6 Canada}
\author{Chapin S. Korosec}
\email{ckorosec@sfu.ca}
\affiliation[SFU]
{Department of Physics, Simon Fraser University, 8888 University Drive, Burnaby, British Columbia, V5A 1S6 Canada}

\title[An \textsf{achemso} demo]
  {Multivalent diffusive transport}

\abbreviations{IR,NMR,UV}
\keywords{}

\begin{document}
\begin{abstract}

We present here a model for multivalent diffusive transport whereby a central point-like hub is coupled to multiple feet, which bind to complementary sites on a two-dimensional landscape. The available number of binding interactions is dependent on the number of feet (multivalency), and on their allowed distance from the central hub (span). Using Monte Carlo simulations that implement the Gillespie algorithm, we simulate multivalent diffusive transport processes for 100 distinct walker designs. {\stkout{Guided}}{\new{Informed}} by our simulation results we derive an analytical expression for the diffusion coefficient of a general multivalent diffusive process as a function of multivalency, span, and dissociation constant $K_\mathrm{{d}}$. {\stkout{This work offers insights into design criteria for multivalent diffusive transport, whereby both the diffusion coefficient and processivity of the system are optimized.}}{\new{Our findings can be used to guide experimental design of multivalent transporters, in particular providing insight into how to overcome trade-offs between diffusivity and processivity.}}   

\end{abstract}
\section{Introduction}

Multivalent interactions are broadly characterized by a central hub that presents multiple copies of a recognition element to a substrate surface displaying complementary binding sites~\cite{Mulder2004, Kiessling2006a}. Such interactions are ubiquitous throughout biology~\cite{Mammen1998, DiIorio2020}. Examples include the adhesion of influenza virus to its host cell achieved through the binding of many hemagglutinin and neuraminidase proteins to sialic acid~\cite{Sakai2017, Muller2019}, and the adhesion of \textit{E. coli} to cell surfaces achieved through multiple pilus interactions~\cite{Mammen1998}. Inspired by Nature, researchers have developed a variety of synthetic multivalent systems~\cite{Mulder2004}. Of relevance to this work are those that walk or diffuse across their substrate landscape~\cite{VonDelius2011, Olah2011, Du2021, Li2018, Pei2006, Lund2010, Yehl2015, Bazrafshan2020a}.

Motivated to design a molecular system that could diffuse across a landscape and release a product, Pei \textit{et al.} designed polycatalytic assemblies, or \textit{molecular spiders}, consisting of a central hub presenting 2 to 6 deoxyribozyme recognition elements that navigate a substrate-decorated landscape via Watson-Crick basepair formation and subsequent substrate cleavage~\cite{Pei2006}. By arranging nucleic acid substrates in distinct 2D patterns, molecular spiders can be controlled to carry out specific sequences of actions~\cite{Lund2010}. They have also been proposed to be `motor-like' and capable of pulling a load~\cite{Samii2010, Olah2013, Morozov2007}, albeit functioning with very low efficiency compared to biological motors~\cite{Samii2011}. Multivalent DNA-based molecular walkers have been designed to carry out a variety of tasks. For example, they have achieved nanoparticle transport~\cite{Cha2014}, operation within living cells whereby DNA motor dynamics are regulated by intracellular interactions~\cite{Peng2017}, and nearly ballistic movement on a two-dimensional plane~\cite{Yehl2015,Bazrafshan2020a}. Despite the multivalent DNA walker field having been active for over 15 years, missing from the literature is a simple study of how much changing the multivalency, as well as the reach of each recognition element from the central hub, over a broad range of values, alters their dynamics. Such insight is crucial towards optimising a system that controls the rate of diffusion of molecules across their landscape as well as to build intuition about their processivity.

\section{Model and methods}
\subsection{Kinetic model}

\begin{figure}[h!]
	\centering
	\includegraphics[width = 1.0\textwidth]{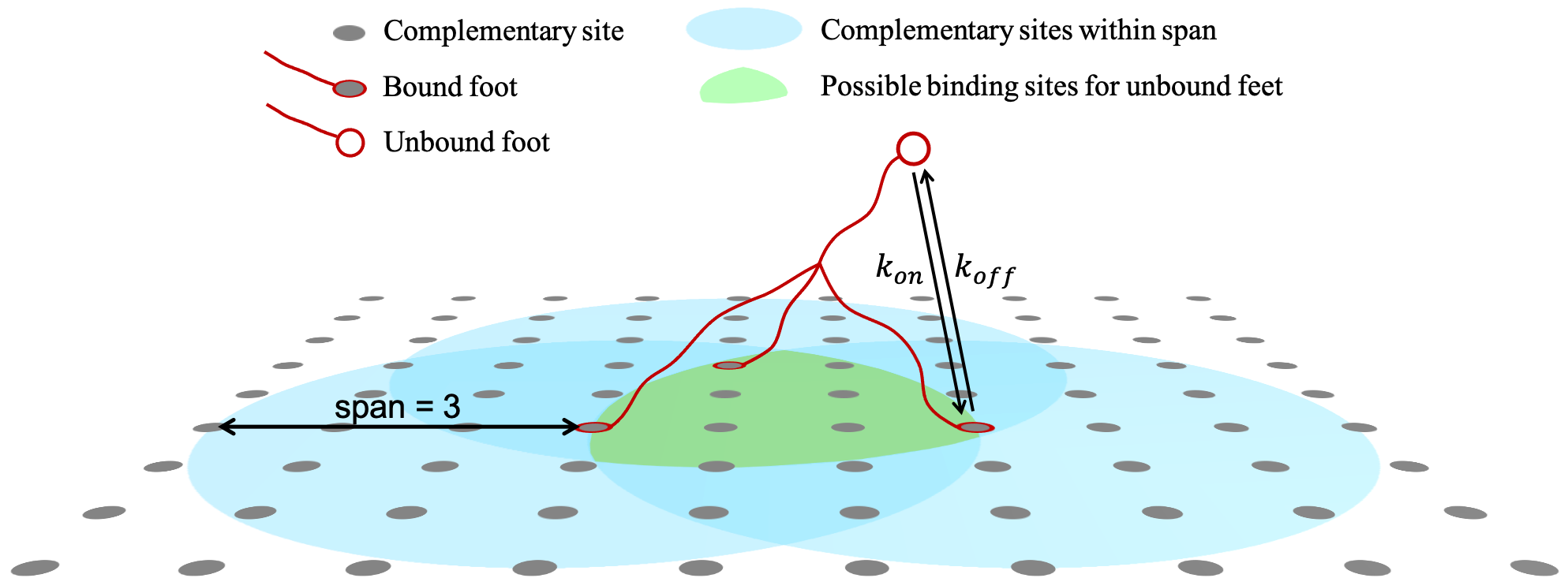}
	\caption[Schematic of a multivalent diffusive walker]{Schematic of a multivalent diffusive walker. Each bound foot is assigned a span which defines the radius of a circle, shown in blue. The mutual overlap of the circles (shown in green) marks the feasible binding locations for unbound feet. Each foot can bind an available complementary site via an on rate $k_{\rm{on}}$, and detach with rate $k_{\rm{off}}$.}
	\label{fig:CH4fig1}
\end{figure}

In this work we adapt a previous model~\cite{Korosec2018} to explore the dynamics of multivalent diffusive transport through Monte Carlo simulations that implement the Gillespie algorithm {\new{(Supporting Information)}}~\cite{Gillespie1977}. A representative schematic of our system is shown in Fig.~\ref{fig:CH4fig1}. We model non-catalytic multivalent walkers with $n$ multivalency and $s$ span, where span defines the distance{\new{ between a bound foot and its farthest reachable complementary binding site.}} {\stkout{from the central hub.}} {\new{This span is then used to define the radius of a circle around each bound foot on the two-dimensional landscape. Unoccupied complementary sites that fall within the mutual overlap of all circles derived from bound feet represent the accessible binding sites for unbound feet. In this work, the span of all feet is always equal for a particular multivalency.}} We explore the parameter space of $(n, s)$ from $(n = 2, s = 2)$ up to $(n = 16, s = 16)$; 100 distinct multivalent walker designs are simulated. Each of the $n$ feet can interact with complementary binding sites on a two-dimensional landscape through a binding rate $k_{\rm{on}}$ and unbinding rate $k_{\rm{off}}$. {\new{$k_{\rm{on}}$ and $k_{\rm{off}}$ are binding rates per accessible site. For example, if there are 10 accessible binding sites and 1 unbound foot, the unbound foot may bind each of the 10 accessible sites with rate $k_{\rm{on}}$.}} Unless otherwise stated, $k_{\rm{on}}$ and $k_{\rm{off}}$ are fixed at 20 and 1, respectively. These rates are chosen such that $k_{\rm{on}} > k_{\rm{off}}$ to ensure the diffusive walkers remain processive (do not detach) for enough time to collect sufficient statistics to reliably compute a diffusion coefficient. No simultaneous binding of multiple feet to a single complementary site is allowed. Each multivalent walker is initialized at the center of a two-dimensional landscape with a single foot bound. A single trajectory lasts until all feet unbind or until 25,000 timesteps of simulation time have elapsed. 

{\new{Our model is intentionally simple. We assume substrate lattice sites to be uniformly arranged on a square lattice with lattice constant defined to be 1. We consider only the slowest timescales in the system: binding/unbinding are generally orders of magnitude slower than diffusion over the relevant lengthscales. We assume independent foot-binding kinetics; thus, implicitly we ignore the effect of tension-induced forces that may enhance $k_{\mathrm{off}}$. This assumption is in line with the expectation that thermally accessible conformations are responsible for binding within the allowed span, and thus are unlikely to generate significant tension. While these assumptions are based on previous work modelling DNA-based walkers~\cite{Samii2010, Samii2011}, they are likely to hold true for other multivalent systems whose rates are limited by binding kinetics.}} 

\subsection{Mean squared displacement}
We are interested in how multivalency and span alter the dynamics of multivalent diffusive transport. For each multivalency and span combination we simulate $N = 500$ trajectories for walkers that remain processive and $N = 20,000$ trajectories for non-processive walkers. The $N$ trajectories are then analyzed by an ensemble trajectory-averaged mean squared displacement (MSD$_\mathrm{ETA}$) given by \begin{equation}\label{eqn:ETAMSD}
    \mathrm{MSD_{\rm ETA}}(\tau) \equiv \left\langle\overline{\Delta r^{2} \left( t, \tau \right) }\right\rangle   = \frac{1}{N}\sum_{j = 1}^{N} \overline{\Delta r_{j}^{2} \left( t, \tau \right)} = \frac{1}{N}\sum_{j = 1}^{N} \left[ \frac{\Delta t}{T_{j} - \tau + \Delta t}  \sum_{t = 0}^{T_{j} - \tau}\Delta r_{j}^{2} \left( t, \tau \right)\right].
\end{equation}
$\Delta r_{j}^{2}$ is the squared displacement {\stkout{for}}{\new{of the walker in}} the $jth$ trajectory at time-lag $\tau$, $T_{j}$ is the duration of the $jth$ trajectory, and $\Delta t$ is the simulation timestep~\cite{Korosec2020a}. {\new{The position of a multivalent walker is defined as the mean position of all bound feet, and is used to compute the walker's displacement, $\Delta r_{j}$.}} For a conventionally diffusive system Eq.\ref{eqn:ETAMSD} can be related to the diffusion coefficient $D$ 
\begin{equation}\label{eqn:diff}
\mathrm{MSD_{\rm ETA}} = 2dD\tau, 
\end{equation}
where $d$ is the dimension of the diffusive process and is equal to 2 for this work. For an anomalously diffusive system the $\mathrm{MSD}$ is represented by
\begin{equation}\label{eqn:MSDanom}
\mathrm{MSD_{\rm ETA}} = K\tau^{\alpha}, 
\end{equation}
where $K$ is the generalized diffusion coefficient and $\alpha$ the anomalous diffusion exponent. For values of $\alpha < 1$, $\alpha = 1$, and $1 < \alpha < 2$, the system is subdiffusive, conventionally diffusive, and superdiffusive, respectively~\cite{Metzler2000}. 
When presenting diffusion coefficients derived from these simulated trajectories, we use the symbol $D_\mathrm{{sim}}$, to indicate that this diffusion coefficient is determined through an MSD analysis completed on the simulated data. 

\section{Results}

\begin{figure}[h!]
	\centering
	\includegraphics[width = 1.0\textwidth]{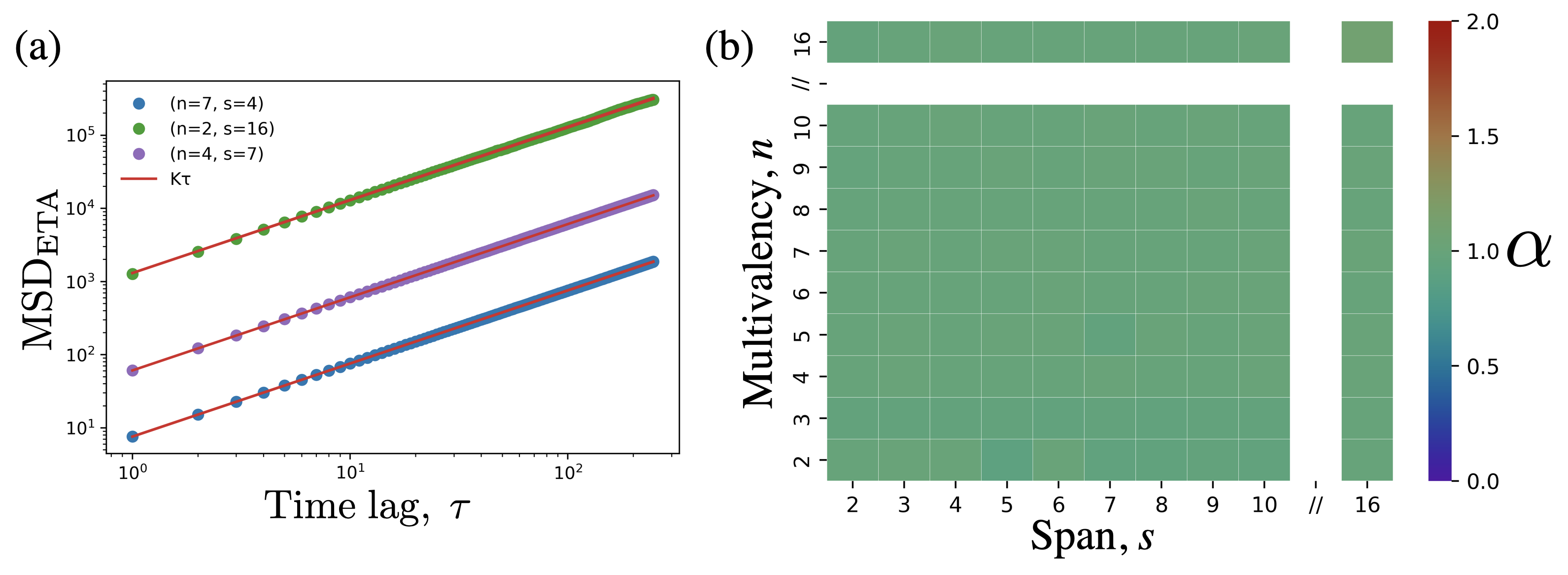}
	\caption[ETA MSD analysis]{$\mathrm{ETA_{MSD}}$ analysis. a) Example linear fits of log-log $\mathrm{MSD_{ETA}}$ versus time lag for various multivalent walker designs. These all have a slope of $\alpha = 1$ (Eq.~\ref{eqn:MSDanom}). b) The anomalous diffusion exponent $\alpha$ is given by the slope of the MSD versus time lag curve and is plotted for all multivalent diffusive transporters; all $\alpha$ values are approximately 1.0, characteristic of conventional diffusion.}
	\label{sfig:CH4sfig1}
\end{figure}

We first validate the use of Eq.~\ref{eqn:diff}{\new{, which assumes the multivalent walkers to undergo conventional difusion}}.  Fig.~\ref{sfig:CH4sfig1}a displays a few examples of $\rm{MSD_{ETA}}$ curves to demonstrate a linear relationship, and along with Fig.~\ref{sfig:CH4sfig1}b which shows that for all $(n,s)$ the system is conventionally diffusive (no anomalous diffusive effects are present), justifies our use of Eq.~\ref{eqn:diff} to accurately determine a diffusion coefficient for our system. 

\begin{figure}[h!]
	\centering
	\includegraphics[width = 0.9\textwidth]{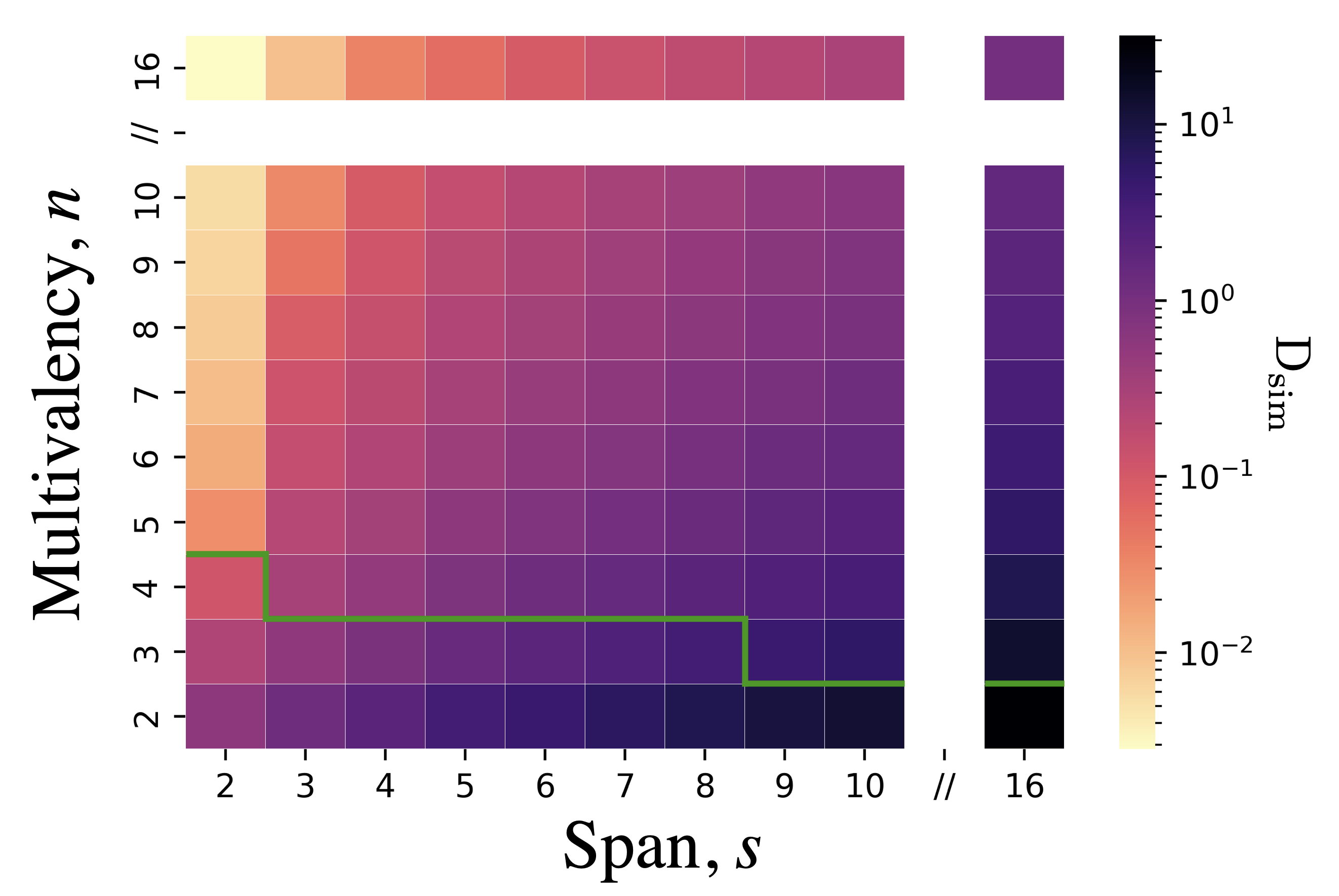}
	\caption[Heatmap of diffusion coefficients]{Heatmap of diffusion coefficient $D_{\mathrm{sim}}$ as a function of multivalency and span. $D_{\mathrm{sim}}$ is minimized for large multivalency and small span and conversely maximized for small multivalency and large span. For a constant multivalency, increasing span leads to an increase in $D_{\mathrm{sim}}$. For a constant span, increasing multivalency leads to a decrease in $D_{\mathrm{sim}}$. The green line indicates a processivity boundary, above which $>99\%$ of walkers of each (n,s) design remained engaged with the landscape throughout the simulation time.}
	\label{fig:CH4fig2}
\end{figure}

Fig.~\ref{fig:CH4fig2} displays a heatmap of the diffusion coefficient, $D_{\mathrm{sim}}$, as a function of multivalency and span, where $D_{\mathrm{sim}}$ is obtained from~Eq.~\ref{eqn:diff}. The green line in Fig.~\ref{fig:CH4fig2} outlines the boundary above which $>99\%$ of walkers of each $(n,s)$ design remain processive within our simulation time of 25,000 simulated timesteps. From Fig.~\ref{fig:CH4fig2}, we see that the diffusion coefficient can be tuned by at least 4 orders of magnitude, keeping the same binding kinetics, simply by varying multivalency and span. For example, $(n = 16, s = 2)$ leads to $D_{\mathrm{sim}} = 0.003$, while $(n = 2, s = 16)$ leads to $D_{\mathrm{sim}} = 31.8$. For any fixed value of multivalency, increasing span results in an increase in $D_{\mathrm{sim}}$. Conversely, for any fixed value of span, increasing multivalency results in a decrease in $D_{\mathrm{sim}}$.

\begin{figure}[h!]
	\centering
	\includegraphics[width = 0.9\textwidth]{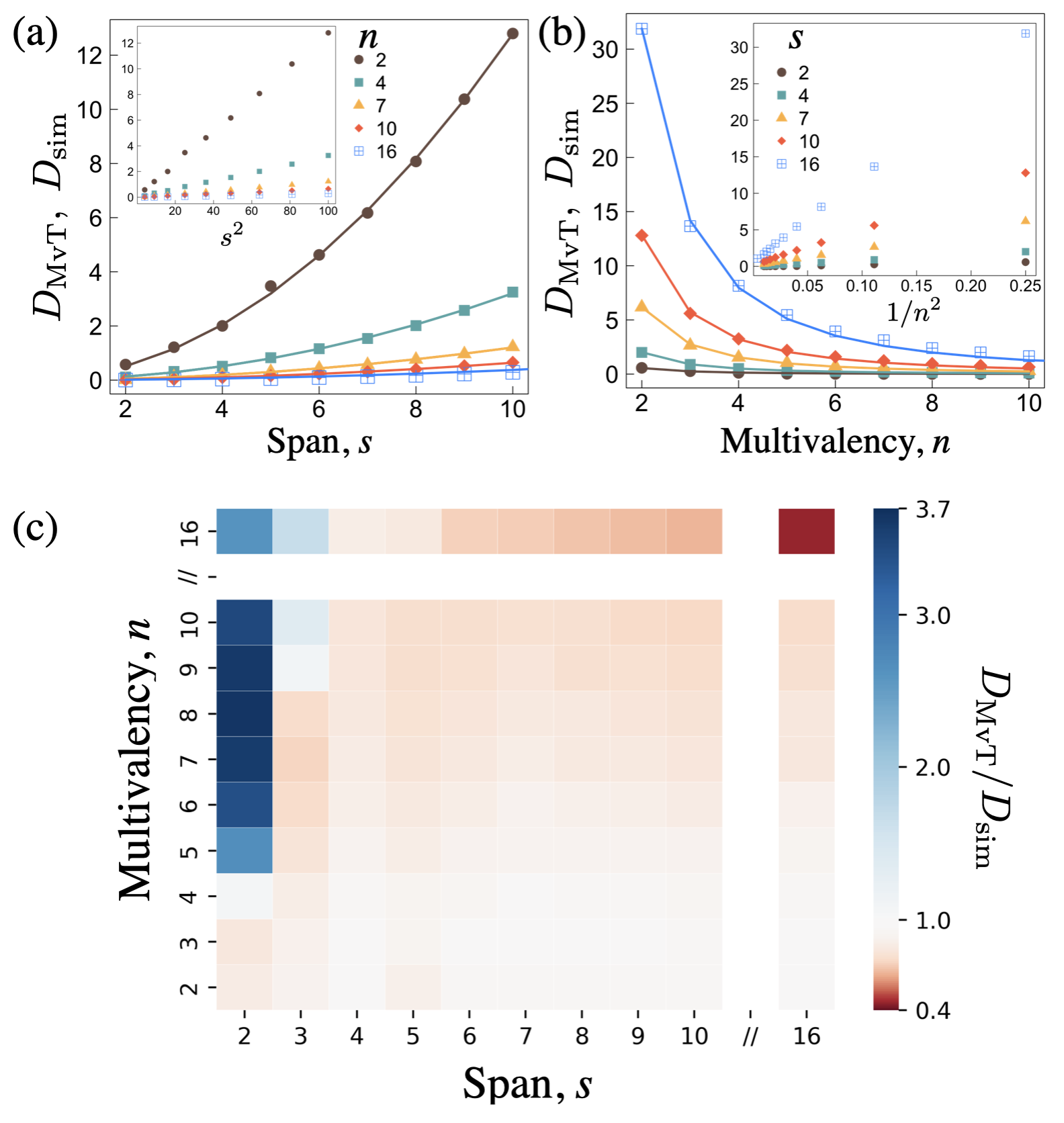}
	\caption[Multivalent walker dynamics and diffusion comparison]{Multivalent walker dynamics and diffusion comparison. For both a) and b) $D_{\mathrm{sim}}$ is shown with points while $D_{\mathrm{MvT}}$ (Eq.~\ref{eqn:Dsnk}) is shown with curves (not a fit). a) $D$ as a function of span for various values of multivalency. Inset displays $D_{\mathrm{sim}}$ as a function of $s^{2}$. b) $D$ as a function of multivalency for various values of span. Inset displays $D_{\mathrm{sim}}$ as a function of $1/n^{2}$. (c) Heatmap of the ratio of the analytically determined diffusion coefficient $D_{\mathrm{MvT}}$ from Eq.~\ref{eqn:Dsnk} to the simulation-determined diffusion coefficient $D_{\mathrm{sim}}$ from Eq.~\ref{eqn:diff}, as a function of multivalency and span. 
	}
	\label{fig:CH4fig3}
\end{figure}

In Fig.~\ref{fig:CH4fig3}a, we show how the diffusion coefficient increases as a function of span. From our model schematic (Fig.~\ref{fig:CH4fig1}) it can be seen that span defines circles of complementary sites around each foot, which overlap to define an area
of accessible binding sites on the two-dimensional landscape. Therefore, as span increases the accessible landscape increases proportionally to $s^{2}$. Thus, it follows that the diffusion coefficient (which has units of length$^{2}$/time) ought to depend proportionally on $s^{2}$. In Fig.~\ref{fig:CH4fig3}a we see indeed that the diffusion coefficient increases as $s^{2}$. 

In Fig.~\ref{fig:CH4fig3}b, we show how the diffusion coefficient decreases as a function of multivalency. The region of accessible binding sites for unbound feet is constrained by all bound feet, and thus as the number of bound feet increases this region becomes more limited. Therefore, as multivalency increases, on average each step of the multivalent walker becomes closer to the central hub leading to a decrease in the diffusion coefficient. {\new{The diffusion coefficient is proportional to the square of the step size, $\ell$, where $\ell$ represents the shift in the center of mass due to the binding of a single foot. We expect that a single foot binding a distance $r$ from the center of mass will shift the center of mass by $r/n$. This suggests that the average step size is proportional to $1/n$, and therefore that the diffusion coefficient is proportional to $1/n^{2}$}}.  {\stkout{Furthermore, as multivalency increases more feet are required to simultaneously unbind and then rebind in new locations in order to substantively move the central hub, which also leads to a slowing of the diffusive process.}} {\new{Indeed, as}}{\stkout{As}} shown in Fig.~\ref{fig:CH4fig3}b, we find that the diffusion coefficient decreases with increased multivalency as $1/n^{2}$. 

\subsection{Multivalent diffusion coefficient derivation}\label{section:DiffDerivation}

The diffusion coefficient for a random walk with discrete time, discrete space, and constant step size can be written as
\begin{equation}
D = \frac{\ell^{2}}{2\Delta t}, 
\end{equation}
where $\ell$ is the step size and $\Delta t$ the time step. We find that for multivalent transport the diffusion coefficient $D$ depends proportionally on $n$ and $s$ as $\frac{1}{n^{2}}$ (Fig.~\ref{fig:CH4fig3}b, inset) and $s^{2}$ (Fig.~\ref{fig:CH4fig3}a, inset), respectively. Thus, we assume an average stepsize $\ell = s/n$ for our system.

The effective time per step in our system is dictated by the foot binding and unbinding times, and thus can be written as
\begin{equation}\label{eqn:timeStep}
\Delta t = \tau_{\mathrm{on}} + \tau_{\mathrm{off}} = \frac{1}{k_{\mathrm{on}}} + \frac{1}{k_{\mathrm{off}}} = \frac{k_{\mathrm{on}} + k_{\mathrm{off}}}{k_\mathrm{on}k_{\mathrm{off}}}. 
\end{equation}
For diffusive multivalent transport we then find the diffusion coefficient to be  
\begin{equation}\label{eqn:DmvtDerivation}
D_{\mathrm{MvT}} =  \frac{1}{2} \frac{k_{\mathrm{on}} k_{\mathrm{off}}}{k_{\mathrm{on}} + k_{\mathrm{off}}} \frac{s^{2}}{n^{2}}, 
\end{equation}
where the subscript MvT denotes `multivalent transport'. We note that Eq.~\ref{eqn:DmvtDerivation} has the required units of a diffusion coefficient of length$^{2}$/time. 

The thermodynamic binding strength of a system with reversible binding interactions can be characterized by its dissociation constant, ${K}_{\rm{d}}$. ${K}_{\rm{d}}$ can be expressed in terms of the unbinding and binding rate constants as ${K}_{\rm{d}} = \frac{k_{\mathrm{off}}}{k_{\mathrm{on}}}$. Eq.~\ref{eqn:DmvtDerivation} can then be rewritten as
\begin{equation}\label{eqn:Dsnk}
D_{\mathrm{MvT}} =  \frac{1}{2} \frac{k_{\mathrm{off}}}{\mathrm{K}_{d} + 1} \frac{s^{2}}{n^{2}}.
\end{equation}
Using Eq.~\ref{eqn:diff} the MSD for multivalent transport in two dimensions can be expressed as
\begin{equation}\label{eqn:DsnKD}
\mathrm{MSD} =  \frac{2k_{\mathrm{off}} s^{2} \tau}{{(K}_{d} + 1)n^{2}}.
\end{equation}
For a strong-binding system (${K}_{\rm{d}}<<1$), Eq.~\ref{eqn:DsnKD} reduces to
\begin{equation}\label{eqn:Dkdred}
\mathrm{MSD} =  \frac{2{k_{\mathrm{off}}} \tau s^{2}}{n^{2}}.
\end{equation}
Therefore for a system where the rate of binding to a given recognition site is much faster than the rate of dissociation, the diffusion constant should be linearly proportional to the off rate. 

\subsection{Comparing $D_{\mathrm{MvT}}$ to $D_{\mathrm{sim}}$}


In Fig.~\ref{fig:CH4fig3}, we show comparisons between $D_{\mathrm{MvT}}$ (Eq.~\ref{eqn:Dsnk}) and our simulation-derived diffusion coefficients, $D_{\mathrm{sim}}$.  We find excellent agreement between these values across almost all of the studied parameter space.  Two exceptions are for walkers with the highest multivalency and span $(n=16, s=16)$ and those with very short span and high multivalency $(n \geq 5, s = 2)$. We suspect that in the limit of large multivalency and low span the assumption that the average step length is equal to $s/n$ is less accurate because there is a saturation of binding sites: it is not possible for all feet to simultaneously bind. In this limit, $D_{\rm MvT}$ is expected to overestimate $D_{\rm sim}$ as it does not take into account the saturation of available binding sites. As $n$ and $s$ increase $D_{\rm MvT}$ appears to increasingly underestimate $D_{\rm sim}$. This is most easily seen in Fig.~\ref{fig:CH4fig3}c at point $(n=16, s=16)$ where the ratio of $D_{\rm MvT}$ to $D_{\rm sim}$ is approximately 0.4. We suspect that this is because our assumptions of step size ($\ell = s/n$) and time step (Eq.~\ref{eqn:timeStep}) do not capture all the rich system details of the Gillespie simulation. For example, the time to the next move in the Gillespie simulation is given by
\begin{equation}\label{eqn:GillespieSim}
    \tau_{\mathrm{G}} = \frac{1}{\lambda_{t}}\mathrm{ln}\left( \frac{1}{r} \right),
\end{equation}
where $\lambda_{t}$ is the total system rate~\cite{Gillespie1977}, and $r$ is a random number drawn from a uniform distribution bounded by $(0,1)$.
{\stkout{$\tau_{\mathrm{G}}$ is dependent on every option that the walker has access to in its current state: s}} {\new{$\lambda_{t}$ depends on the number of bound and unbound feet, and on the number of accessible binding sites, which in turn depends on the number, location, and span of bound feet. Therefore, as $n$ and $s$ increase, the average number of transitions contributing to $\lambda_{t}$ increases nonlinearly, and stochastically varies throughout a simulation. S}}uch details are not incorporated in Eq.~\ref{eqn:timeStep}{\new{, which provides a simple estimate of the time required for the binding/unbinding cycle for each individual foot}}. 
{\stkout{Furthermore, while $\ell = s/n$ is a simple conceptual estimate of the step size, the true power dependence of $\ell$ on $s$ and $n$ might deviate slightly from 1 and -1, respectively.}} Further work on analytically determining a more accurate timestep or step length is beyond the scope of this work. 

\begin{figure}[h!]
	\centering
	\includegraphics[width = 0.6\textwidth]{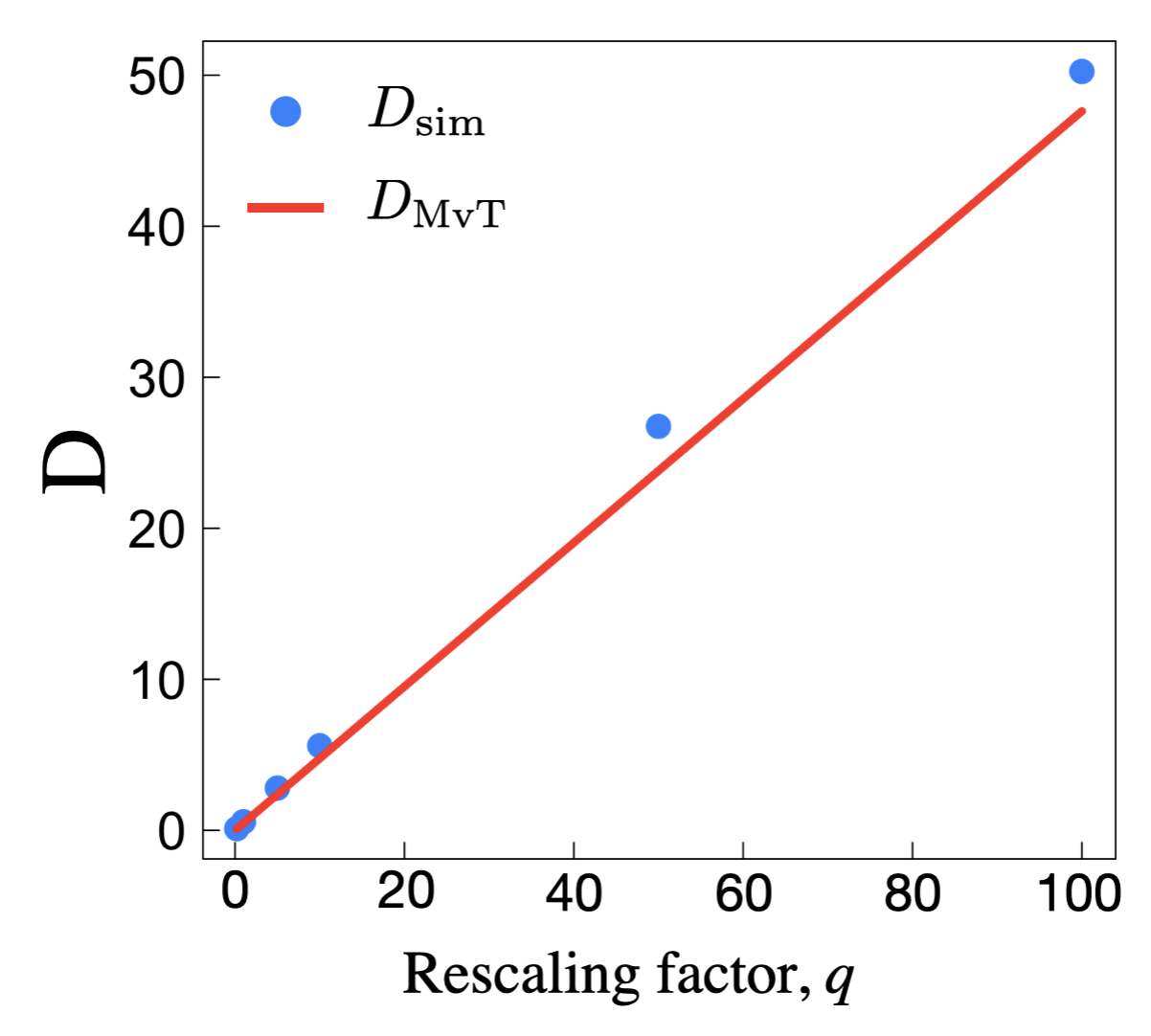}
	\caption[$D$ as a function of $q$ for $n = s = 5$]{$D$ as a function of the kinetic rescaling factor $q$ for $n = s = 5$. Simulated results (blue circles) are slightly higher than estimated results from Eq.~\ref{eqn:Dsnk} (red line).}
	\label{sfig:CH4sfigNew1}
\end{figure}

In addition to multivalency and span we are also interested in how the kinetic rates $k_{\mathrm{on}}$ and $k_{\mathrm{off}}$ alter the dynamics of multivalent transport. In experimental DNA-based walkers $k_{\mathrm{off}}$ can be controlled by the number of basepairs in the nucleic acid duplex~\cite{Pei2006}. 
How closely matched are $D_{\mathrm{sim}}$ and $D_{\mathrm{MvT}}$ for varying values of $k_{\mathrm{on}}$ and $k_{\mathrm{off}}$? In Fig.~\ref{sfig:CH4sfigNew1} $K_{\mathrm{d}}$ is fixed while $k_\mathrm{on}$ and $k_\mathrm{off}$ are rescaled by a factor $q$, where $q$ ranges from 0.2 to 100. We find that~Eq.~\ref{eqn:Dsnk} captures the overall trend from the simulations (Fig.~\ref{sfig:CH4sfigNew1}). 

\subsection{Processivity analysis}

As researchers seek to engineer systems that control the rate of diffusion of molecules across their landscape, the ability of the molecular walker to remain processive is a key consideration. Here, we define processivity as the amount of time a walker spends associated with its landscape before it detaches. 

 \begin{figure}[h!]
 	\centering
 	\includegraphics[width = 1.0\textwidth]{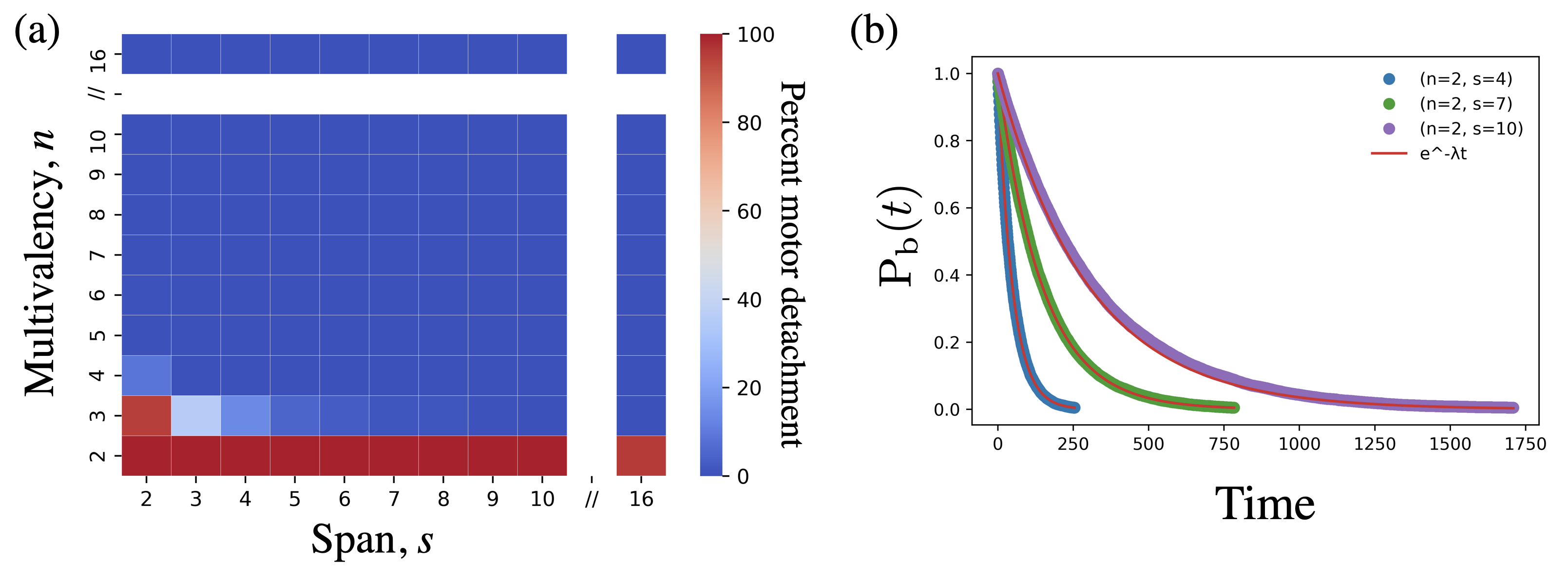}
 	\caption[Processivity analysis]{Processivity analysis. a) {\new{Heatmap depicting the fraction of walkers that detached over the 25000 simulation timesteps. b)}} Example plots of the probability that walkers remain bound to the track versus time for various bipedal walker designs. All $P_{\mathrm{b}} (t)$ curves are described by a single-exponential decay (Eq.~\ref{eq:detachment}).{\stkout{b) Heatmap depicting the fraction of walkers that detached over the 25000 simulation timesteps.}}}
 	\label{sfig:CH4sfig2}
 \end{figure}

In Fig.~\ref{fig:CH4fig2} the green line indicates the boundary above which more than 99$\%$ of multivalent walkers of each $(n,s)$ design class remained associated for the entirety of the simulation time window (25,000 simulated timesteps). In Fig.~\ref{sfig:CH4sfig2}, {\new{a heatmap to visualize processivity across all $(n,s)$ combinations explored in this work is provided, as well as}} example curves of probability of remaining bound ($\mathrm{P}_{\mathrm{b}}(t)$) as a function of time for particular $(n,s)$ combinations{\stkout{are provided, as well as a heatmap to visualize processivity across all $(n,s)$ combinations explored in this work}}. {\new{Overall we find that decreasing multivalency and/or span leads to quicker detachment of multivalent walkers from their landscape. Conceptually this makes sense: as multivalency is decreased it is more probable that all feet simultaneously detach, and therefore the time to complete detachment is expected to decrease. Furthermore, as span is decreased (at fixed multivalency) less of the landscape becomes available for binding, leading to fewer accessible binding sites. The expected time to complete detachment is then also expected to decrease. Indeed, decreased multivalency and span have been linked to less processive molecular spider and burnt-bridge ratchets~\cite{Pei2006, Samii2011, Korosec2018, Korosec2020RM}.}}

\begin{figure}[h!]
 	\centering
 	\includegraphics[width = 1.0\textwidth]{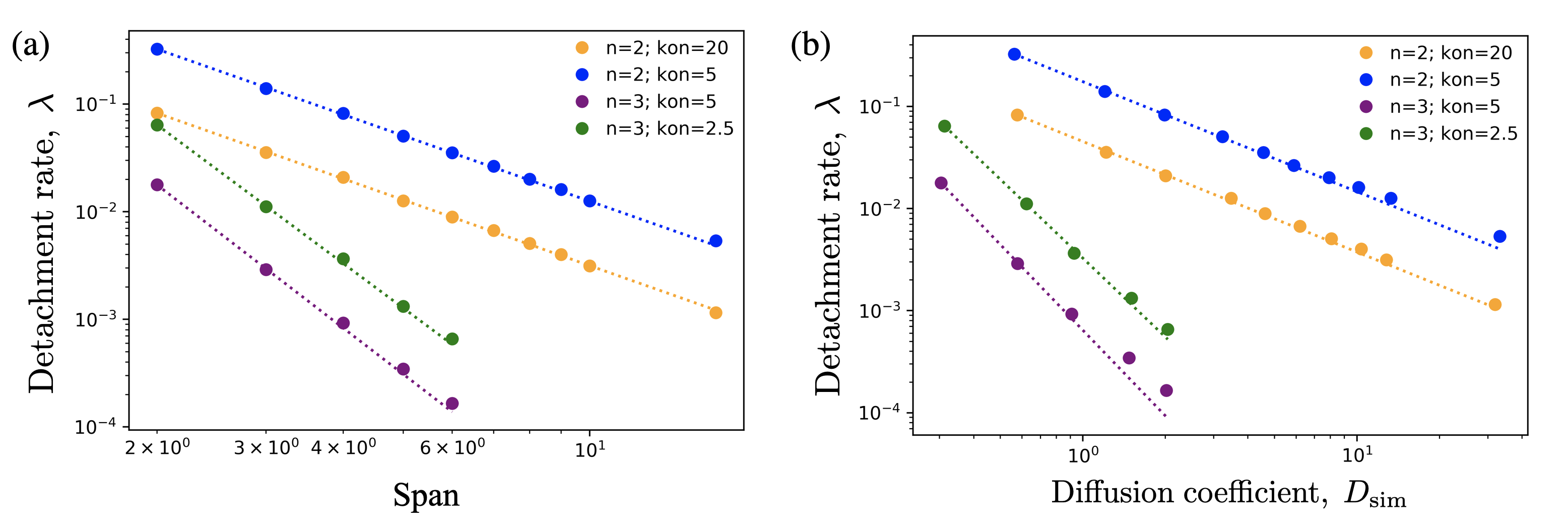}
 	\caption[Detachment rates]{\new{Trade-off between the diffusion coefficient, $D_{\mathrm{sim}}$, and the detachment rate, $\lambda$. $k_{\mathrm{off}} = 1$ in all cases. a) Simulation output for various values of multivalency and $k_{\mathrm{on}}$ plotted in the space of $\lambda$ and $D_{\mathrm{sim}}$. Span increases from left to right along each coloured line. b) $\lambda$ as a function of span for various values of multivalency and $k_{\mathrm{on}}$. Dotted lines represent fits to Eq.~\eqref{eq:lambda}.}}
 	\label{sfig:new_DLambda}
 \end{figure}

For all multivalent walker designs that exhibit significant detachment, we find exponential detachment kinetics described by
\begin{equation}\label{eq:detachment}
\mathrm{P}_{\mathrm{b}}(t) =\rm{e}^{-\lambda t}{\stkout{.}}{\new{,}}
\end{equation}
{\stkout{Here,}}{\new{where}} $\lambda$ is a constant that characterizes the rate of detachment{\stkout{, and varies with walker design}}. {\new{From fits to Eq.~\ref{eq:detachment}, we determined that the detachment rate for fixed $n$ can be described by 
\begin{equation}\label{eq:lambda}
\mathrm{\lambda}(s) = {a}{s^{-b}},
\end{equation}
where $a$ and $b$ are constants (Fig.~\ref{sfig:new_DLambda}a). For example, for $n=2$, we find $a=\frac{1}{3}$ and $b=2$ (Supporting Figure 1). We anticipate $a$ and $b$ to depend on $n$, $k_{\mathrm{on}}$, and $k_{\mathrm{off}}$, as all of these design parameters will affect the average association time of a multivalent walker.}}

{\new{To elucidate potential trade-offs between diffusion coefficient and processivity, we compared $D_{\mathrm{sim}}$ and $\lambda$ parametrically (Fig.~\ref{sfig:new_DLambda}b). The results show that these two system properties can be tuned independently via system design parameters ($i.e.$, $s$, $n$, $k_{\mathrm{on}}$, and $k_{\mathrm{off}}$). For example, a diffusion coefficient of $\approx1$ can be achieved with different $n=2$ and $n=3$ designs, which exhibit distinct detachment rates $\lambda$. Generally a low detachment rate is preferred, corresponding to high processivity; for the bipedal and tripedal systems shown in (Fig.~\ref{sfig:new_DLambda}b), this is achieved for high $k_{\mathrm{on}}$, large span and greater multivalency. Further improvements in processivity can be achieved by increasing multivalency (Fig.~\ref{sfig:CH4sfig2}a), though tuning $n$ alone results in a decreased diffusion coefficient (Fig.~\ref{fig:CH4fig2}), which can be compensated for by increasing span $s$ or by increasing $k_{\mathrm{on}}$.}}

{\stkout{Overall we find that decreasing multivalency and/or span leads to quicker detachment of multivalent walkers from their landscape. Conceptually this makes sense: as multivalency is decreased it is more probable that all feet simultaneously detach, and therefore the time to complete detachment is expected to decrease. Furthermore, as span is decreased (keeping multivalency fixed) less of the landscape becomes available for binding, leading to fewer accessible binding sites. The expected time to complete detachment is then also expected to decrease. Indeed, decreased multivalency and span have been linked to less processive molecular spider and burnt-bridge ratchets~\cite{Pei2006, Samii2011, Korosec2018, Korosec2020RM}.}}

{\stkout{We find that for $n = 2$ the rate of detachment is equal to $\lambda = 1/3s^{2}$, as determined from a fit to Eq.~\ref{eq:detachment} (Fig.~\ref{sfig:CH4sfignew2}). We suspect $\lambda$ to depend on $n, s, k_{\mathrm{on}}$ and $k_{\mathrm{off}}$, as all of these design parameters will affect the average association time of a multivalent walker; the coefficient of $1/3$ found may vary with all these parameters.}}


\section{Discussion}


In engineering a multivalent transport system that maximizes the diffusion coefficient, based on our results presented in Fig.~\ref{fig:CH4fig2}{\new{,}} one would select a walker design with minimum multivalency and a maximum span. However, these minimally multivalent walkers quickly dissociate from the landscape, and therefore are not useful for covering large distances in a processive manner. {\new{Thus, there is an apparent trade-off between diffusivity and processivity. We have shown that there is sufficient room in ($n,s$) parameter space to design walkers that exhibit excellent performance measures for both diffusion and processivity, for example by selecting for intermediate multivalency and maximal span. We have also shown that further improvements in walker dynamics can be achieved by tuning the foot binding kinetics.}}{\stkout{Taking processivity into account, the optimal multivalent walker contains intermediate multivalency (such as $n = 4$) and maximized span}}.

Our simple analytical expression Eq.~\ref{eqn:Dsnk} for multivalent diffusion is successful at capturing the overall trends in the data for a wide range of multivalency and span values (see Fig.~\ref{fig:CH4fig3}c).  For small values of span ($s = 2$) we suspect that a saturation of binding options is causing a deviation of $D_{\mathrm{MvT}}$ from $D_{\mathrm{sim}}$: when all the binding sites are saturated with `feet' the diffusion coefficient should no longer be significantly affected by a further increase in $n$. This detail is not included in our analytical model.

{\new{Our work may also bear significance to the design of DNA-functionalized nanostructures where the goal is to optimize molecular affinity. Synthetic DNA devices have been developed to tightly bind a receptor, and have found applications for viral sensing~\cite{Kwon2020} as well as for exploring the effect ligand patterning has on avidity~\cite{Deal2020}. Where we have explored the timescale of molecule binding as a function of multivalency and span, our results may inform the future design of novel DNA-functionalized devices.}}


Our multivalent diffusers resemble end-modified star polymers. In the absence of specific binding of their ends to a substrate, the three-dimensional diffusion coefficient in solution for star polymers has been found to vary exponentially with multivalency~\cite{Shull1990}. We hypothesize that star polymers end-modified to interact with a complementary two-dimensional landscape will have a diffusion coefficient well approximated by Eq.~\ref{eqn:Dsnk}. Such a system, to the best of our knowledge, has not thus far been experimentally realized, but would be a candidate system to test our newly derived analytical expression.

\section{Conclusions}

In this work we implemented the Gillespie Monte Carlo method to explore the dynamics of multivalent diffusion transport. We found that changing the span and multivalency of a multivalent walker can alter its diffusion coefficient by many orders of magnitude (Fig.~\ref{fig:CH4fig2}). We further derived an analytical expression for the diffusion coefficient that describes well the simulated diffusion coefficient (Fig.~\ref{fig:CH4fig3}c). We explored the effects of processivity in multivalent walker designs, where we found that {\new{the processivity of a walker can easily be tuned through its design parameters. For instance,}} systems designed with only 2 feet are far less processive than those with $n > 2$ feet. Span also influences processivity
{\new{; we have shown an example of this for systems with $n=3$ feet and $K_{\mathrm{d}} = 1/20$, which}} display exponential detachment for low span values, but become processive ($>99\%$) remain attached within our simulation time window with intermediate span values (Fig.~\ref{sfig:CH4sfig2}). Our results are useful towards understanding engineering design principles of multivalent transport systems.

\begin{acknowledgement}
This work was funded by the Natural Sciences and Engineering Research Council of Canada (NSERC) through Discovery Grants to NRF and a Postgraduate Scholarship--Doctoral to CSK. Computational resources were provided by Compute Canada.
\end{acknowledgement}


\newpage 

\bibliography{library.bib}





\end{document}


\beginsupplement
\maketitle 

\begin{center}
\begin{tabular}{c}

\end{tabular}
\author{Antonia Kowalewski}\\
\author{Nancy R. Forde}\\
\author{Chapin S. Korosec}\\
\author{Corresponding author emails:\\ nforde@sfu.ca \\ chapinSKorosec@gmail.com} 
\end{center}

\thispagestyle{empty}
\newpage



\section{Implementation of the Gillespie algorithm}

We use the Gillespie algorithm~\cite{Gillespie1977} to simulate the kinetics of a system that has access to many different moves, each with its own rate $\lambda_{\mathrm{i}}$.

The total system rate $\lambda_{\mathrm{t}}$ is 
\begin{equation}
\lambda_{\mathrm{t}} = \sum_{i = 1}^{m} \lambda_{\mathrm{i}},
\end{equation}
where the sum is over all $m$ rates indexed by $i$. The accessible moves available to the simulated system are determined by the current system state, and encoded into $\lambda_{\mathrm{t}}$. These transitions include, for each unbound foot, its binding (at rate $k_{\mathrm{on}}$) to each accessible substrate site, and for each bound foot, its unbinding (at rate $k_{\mathrm{off}}$).Therefore, $\lambda_{\mathrm{t}}$ must be updated every timestep. The simulation time is updated by computing 
\begin{equation}\label{eqn:GillespieSim}
\tau_{\mathrm{sim}} = \frac{1}{\lambda_{\mathrm{t}}}\mathrm{ln}\left( \frac{1}{r}\right),
\end{equation}
which gives the time step for that transition. $\tau_{\mathrm{sim}} $ is determined by choosing for $r$ a random variable R$_{1}$ from a uniform distribution bounded by $(0,1)$. 

How do we decide which move to make? This too is determined stochastically, by choosing a second random number $\mathrm{R_{2}}$ drawn from a uniform distribution bounded by $(0,1)$. The probability of transition $i$ occurring is given by its weight relative to the total system rate, $\mathrm{P_{\mathrm{i}}} = \lambda_{\mathrm{i}}/\lambda_{\mathrm{t}}$. An easy way to do this is to store the $m$ accessible rates $\lambda_{\mathrm{i}}$ in a one-dimensional array, and use the product $R_{2}\lambda_{\mathrm{t}}$ to determine which of the $m$ transitions is to be chosen for the next move. The chosen transition for the next move corresponds to the smallest integer $k$ that satisfies
\begin{equation}\label{eqn:GillespieaAlgorithm}
R_{2}\lambda_{\mathrm{t}} < \sum_{i = 1}^{k} \lambda_{\mathrm{i}}. 
\end{equation}
 
\begin{figure}[h!]
 	\centering
 	{\includegraphics[width = 0.7\textwidth]{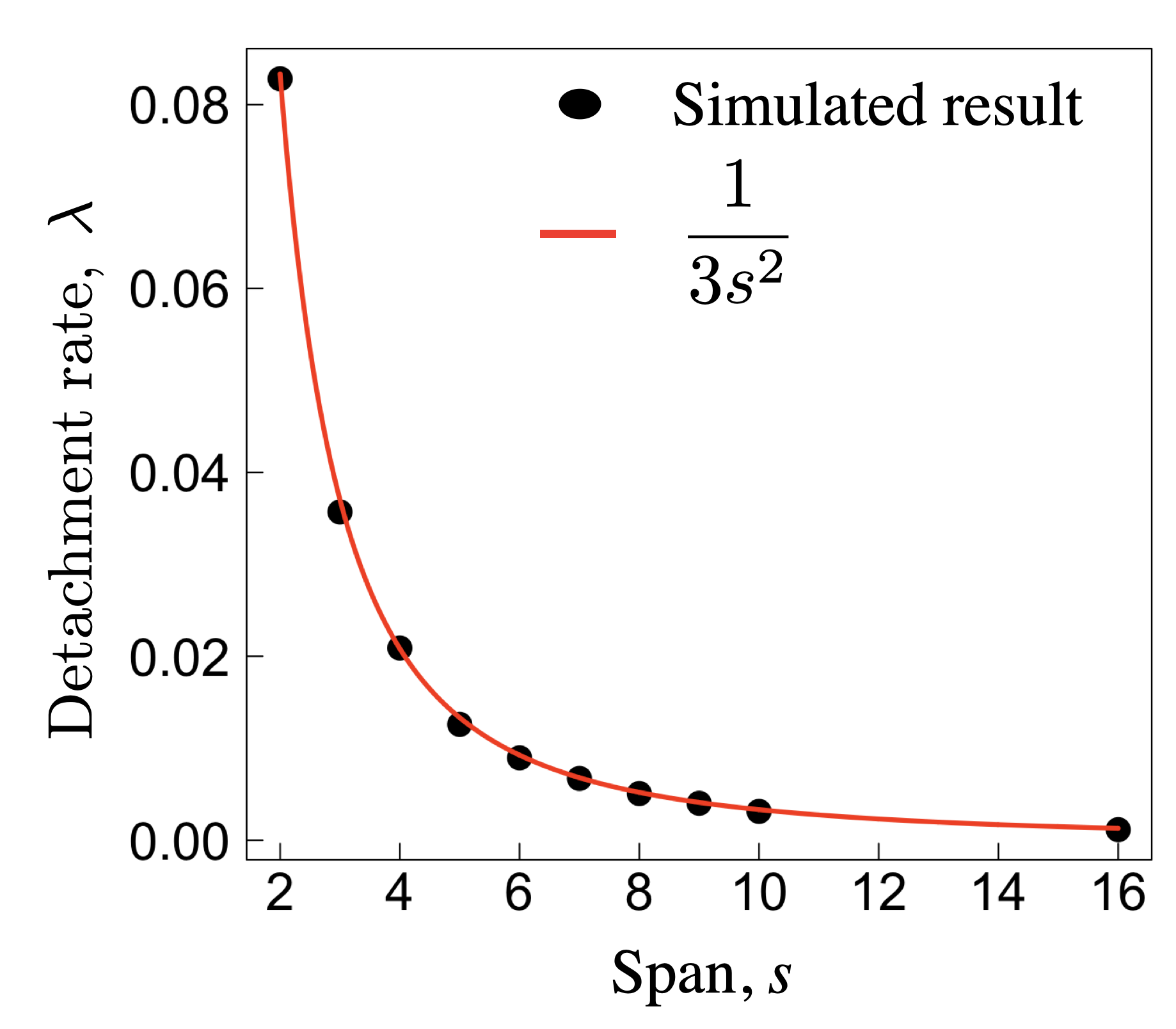}}
 	\caption[Detachment rates]{Detachment rates, $\lambda$, determined from exponential fits of the probability of bound walkers remaining, $\mathrm{P}_{\mathrm{b}}(t)$ (Eq. 11 of the main text), for each $n=2$ multivalent walker design.}
 	\label{sfig:CH4sfignew2}
 \end{figure}
 
\newpage
\bibliographystyle{ieeetr}
\bibliography{library.bib}